\documentclass[letterpaper,12pt]{article}
\usepackage{graphicx}
\usepackage{amsmath}
\usepackage{amsfonts}
\usepackage{hyperref}
\usepackage{amsthm}
 \usepackage{amscd}
\usepackage{mathrsfs}
\usepackage{caption}
\usepackage{subcaption}
\usepackage{float}
\usepackage[top=1in,left=1in,right=1in,bottom=1in]{geometry}

\newtheorem{theorem}{Theorem}[section]

\newtheorem{remark}[theorem]{Remark}

\hypersetup{
	colorlinks = true,
	linkcolor  = red,
	citecolor  = green,
	filecolor  = cyan,
	urlcolor   = magenta,}
\newcommand \bR {\mathbb{R}}

\newcommand{\bCp}{\mathbb{C}^+}

\newcommand {\bCpb} {\overline{\mathbb{C}^+}}

\newcommand{\ds}{\displaystyle}
\newcommand\beq{\begin{equation}}
\newcommand\ene{\end{equation}}

\numberwithin{equation}{section}

\title{Reverse Lieb--Thirring inequality  for the half-line matrix Schr\"odinger operator\thanks{Keywords Spectral inequalities, matrix Schr\"odinger equations, Lieb-Thirring inequalities.} \thanks{AMS 2020 Subject Classifications 34L15; 34L40; 81Q10.}}
\author{
Ricardo Weder\thanks{weder@unam.mx. Emeritus fellow Sistema Nacional de Investigadores, CONAHCYT, M\'exico.}\\
Departamento de F\'\i sica Matem\'atica\\
Instituto de Investigaciones en
Matem\'aticas Aplicadas y en Sistemas\\
Universidad Nacional Aut\'onoma de M\'exico\\
Apartado Postal 20-126, IIMAS-UNAM\\
 Ciudad de M\'exico, CP 01000, M\'exico}

\date{}

\begin{document}
\baselineskip=20 pt

\maketitle

\begin{abstract}
\noindent We prove a reverse Lieb--Thirring inequality with a sharp constant for the matrix Schr\"odinger equation on the half-line.
\end{abstract}

\section{Introduction}
In their celebrated work  Lieb and Thirring \cite{lieb-thi} introduced a family of inequalities that are now known as Lieb--Thirring inequalities. These inequalities, or more precisely the Lieb--Thirring inequalities in spectral form,  are concerned with the negative eigenvalues, $\lambda_j,$  of the one particle Schr\"odinger operator $-\Delta+V.$ They bound the Riesz means $\sum_{j} |\lambda_j|^\gamma$ in terms of $L^p$ norms of the potential  $V.$ Lieb and Thirring  introduced their inequalities in their study of the stability of matter. However, these inequalities have found numerous applications in other problems in functional analysis and mathematical physics. The recent monograph  \cite{flw} contains an extensive study of Lieb--Thirring inequalities. Here we will content ourselves of stating results in one dimension,  that are more closely related to our work. It was proved in \cite{lieb-thi} for $ \gamma > 1/2,$ and by \cite{wei} for $\gamma=1/2,$ that,
\beq\label{1.1}
\sum_j |\lambda _j|^{\gamma} \leq L_{\gamma,1} \int_{\mathbb R} V_-(x)^{1/2+\gamma}\, dx,
\ene
with $V = V_+-V_-,$ where  $V_\pm:= 1/2(|V|\pm V)$ are, respectively, the positive and the negative parts of the potential $V$.  Further, $ V_-\in L^{1/2+\gamma}(\mathbb R),$ and $V_+\in L^1_{\text{\rm loc}}(\mathbb R).$ Moreover, $L_{1,\gamma}$ is a constant independent of $V.$ In the case $\gamma=1/2,$ $L_{1/2,1}=1/2,$ and this value is sharp \cite{hunder}. 
In \cite{elu} Lieb--Thirring inequalities were proved for the matrix Schr\"odinger operator on the half-line. Namely, they consider the selfadjoint Schr\"odinger operator  $-\frac{d^2}{dx^2}+V(x)$ in  
 $L^2(\mathbb R^+, \mathbb C^n),$ for $ n=1,\dots,$  with the  boundary condition $\psi'(0)= B \psi(0),$
where $V(x)$ is a selfadjoint, $ n\times n$ matrix that satisfies, $ V\geq 0,$
$$
\int_0^\infty \text{\rm Tr}[V^2(x)] \, dx < \infty,
$$
and $B$ is a selfadjoint $n \times n$ matrix. 

  One of their main results is the following.They prove that the negative spectrum of the Schr\"odinger operator consists of  eigenvalues $\lambda_j,$ with multiplicity $ m_j.$ Further, the following Lieb--Thirring estimate holds \cite{elu},
\beq\label{1.2}
\frac{3}{4}|\lambda_1| \,\text{\rm Tr}[B]+\frac{1}{2}(2m_1-n)  |\lambda_1|^{3/2} + \sum_{j\geq 2}  m_j|\lambda_j|^{3/2}\leq
\frac{3}{16} \int_0^\infty \text{\rm Tr}[V^2(x)] \, dx+ \frac{1}{4} \text{\rm Tr}[B^3].
\ene
For further results on Lieb--Thirring inequalities on the  half-line in the scalar case see \cite{ef} and \cite{schimm}.

In this paper we are interested in reverse Lieb--Thirring inequalities in one dimension. Namely, in inequalities where one bounds  from below  a Riesz mean of the absolute value of the negative eigenvalues by the integral of the potential. This type of inequality was first proved independently by Glaser et al. \cite{glaser}, and by  Schminke \cite{schmi}. It was proved by these authors that,
\beq\label{1.3}
\sum_j \sqrt{|\lambda_j|} \geq  -\frac{1}{4} \int_{\mathbb R} V(x)dx,
\ene
where the potentials $V$  is integrable. Furthermore, the constant $ 1/4$ is sharp. See also \cite{dr} for a further reverse Lieb--Thirring inequality. Moreover, see \cite{bfs} for two sided Lieb--Thirring inequalities in terms of the landscape function.

The aim of this paper is to prove a reverse Lieb--Thirring inequality for the matrix Schr\"odinger operator on the 
half-line.  For the results below in the half-line matrix Schr\"odinger operator the reader can consult \cite{AW2021}. Let us consider the formal matrix Schr\"odinger operator  in  $L^2(\mathbb R^+, \mathbb C^n),$ for $ n=1,\dots,$

\beq\label{1.4}
-\frac{d^2}{dx^2}+ V(x),
\ene
where the potential $V$ is an $n\times n$ selfadjoint matrix-valued function.
We  assume that the potential $V$ is integrable, i.e. it satisfies
\begin{equation}\label{1.4b}
\int_0^\infty dx\,\,|V(x)|< \infty,
\end{equation}
where $|V(x)|$ denotes the
operator norm of the matrix $V(x).$
We obtain a selfadjoint Schr\"odinger operator on the half line
by supplementing the formal matrix Schr\"odinger operator \eqref{1.4} with the
general selfadjoint boundary condition at $x=0,$
which is written as
\begin{equation}
\label{1.5}
-B^\dagger \psi(0)+A^\dagger \psi'(0)=0,
\end{equation}
where the  $n\times n$ matrices $A$ and $B$ satisfy
\begin{equation}
\label{1.6}
B^\dagger A= A^\dagger B,
\end{equation}
\begin{equation}
\label{1.7}
A^\dagger A+B^\dagger B>0,\end{equation}
and we refer to $A$ and $B$ as the boundary matrices. 
Postmultiplying the boundary matrices on the right by an invertible
$n\times n$ matrix $T$ does not change \eqref{1.5}. Thus, even though the boundary condition \eqref{1.5}  is uniquely
determined by 
the boundary-matrix pair $(A,B),$ 
the matrix pair $(AT,BT)$ with any invertible matrix $T$
also yields the
same boundary condition \eqref{1.5}. Actually, as proved in Proposition 2.2.1 of \cite{AW2021}, this is the only freedom that we have in choosing the matrices $A,B.$ We denote by $H_{A,B}(V)$ the selfadjoint realization in $ L^2(\mathbb R^+, \mathbb C^n)$ of the formal Schr\"odinger operator $-\frac{d^2}{dx^2}+ V(x)$ with the boundary condition \eqref{1.5}, where the boundary matrices $A,B$ satisfy \eqref{1.6}, \eqref{1.7}. For the details in the construction of  $H_{A,B}(V)$ see Sections 3.3, and 3.5 of \cite{AW2021}. As proved in Sections 3.4 and 3.6 of \cite{AW2021}, we can unitarily transform the operator  $H_{A,B}(V)$ into the  operator $H_{\hat{A},\hat{B}}(\hat{V}):= M  H_{A,B}(V) M^\dagger,  \hat{V}:= M V M^\dagger,$ where $M$ is a unitary matrix, and
$A= M \hat{A} T_1 M^\dagger T_2,  B= M \hat{B} T_1 M^\dagger T_2,$ for some invertible matrices $T_1,T_2,$ and where
\beq\label{1.8}
\hat{A}=- \text{\rm diag}\{ \sin \theta_1, \dots, \sin \theta_n\}, \hat{B}= \text{\rm diag}\{ \cos \theta_1, \dots, \cos \theta_n\}, 0 < \theta_j \leq \pi, j=1,\dots,n.
\ene
With the boundary matrices \eqref{1.8} the boundary condition \eqref{1.5} takes the form
\beq\label{1.9}
(\cos\theta_j) \psi_j(0)+(\sin\theta_j) \psi'(0)=0, \qquad j=1,\dots,n.
\ene
By \eqref{1.9}, we  see that in the representation where the boundary matrices are diagonal we have Dirichlet boundary condition when $\theta_j=\pi,$  Neumann boundary condition when $\theta_j= \pi/2,$ and mixed boundary condition if $ \theta_j \neq \pi/2, \pi.$ Further, we have no Dirichlet boundary condition in \eqref{1.9} if and only if the boundary matrix  $A$ is invertible.
 
For our reverse Lieb--Thirring inequality we consider the boundary condition \eqref{1.5} with the boundary matrix $A$ invertible. As mentioned above, this amounts to exclude Dirichlet boundary conditions in the diagonal representation of the boundary matrices. We exclude Dirichlet boundary conditions to obtain a meaningful reverse Lieb--Thirring inequality, as we explain in Remark~  \ref{remark} below. Note that if $A$ is invertible we can take $T= A^{-1}$ and transforming $(A,B)$ into $(A T,  B T)$ we obtain the boundary matrices $(I, B A^{-1}).$ Hence, in the case where $A$ is invertible there is no loss of generality in considering the operator $H_{I,B}(V),$ with the boundary condition
\beq \label{1.10}
\psi'(0)= B \psi(0).
\ene
Observe that for the pair $(I,B)$ conditions \eqref{1.6}, \eqref{1.7} just amount to require that  $B$ is selfadjoint.
It follow from Theorems  3.11.1 and 4.3.3 of \cite{AW2021} that if the potential satisfies \eqref{1.4b} the operator $H_{I,B}(V)$ has no singular continuous spectrum, that its absolutely continuous spectrum is $[0,\infty),$ and that it has no positive eigenvalues. Further, zero can be an eigenvalue, and there are $N$  negative eigenvalues $\lambda_j,$ with multiplicity $ m_j\leq n,$ for $ j=1,\dots.$ The number of negative eigenvalues $N$ can be zero, finite, or infinite. If there are an infinite number of negative eigenvalues they accumulate at zero. Our reverse Lieb--Thirring inequality is given in  the following theorem.
\begin{theorem}\label{theorem1}
Let $H_{I,B}(V)$ be the selfadjoint realization in $L^2(\mathbb R^+, \mathbb C^n)$ of the formal matrix Schr\"odinger operator \eqref{1.4} with the boundary condition \eqref{1.10} where $B$ is a selfadjoint matrix, and the potential $V$ is selfadjoint and   fulfills \eqref{1.4b}. Assume that $H_{I,B}(V)$ has negative eigenvalues  $\lambda_j,$ for $j=1,\dots.$ Then, the following reverse Lieb--Thirring inequality holds,
\beq\label{1.11}
 \sum_{j}   m_j^{} \sqrt{\left|\lambda_j\right|} \ds  > \frac{1}{4}\left [
 - \int_0^\infty \text{\rm Tr}\left[V(x)\right]\, dx - \text{\rm Tr}[B]  \right],
 \ene
where the constant $1/4$ is sharp.
\end{theorem}
  In the scalar case, $n=1,$ Theorem~\ref{theorem1} is given in \cite{bt} assuming that $V$ is integrable and that there is only a finite number of negative eigenvalues, $ \lambda_j,$ for $ j=1,\dots, N < \infty. $ In our Theorem~\ref{theorem1}  the number of negative eigenvalues is allowed to be infinite. The proof of \eqref{1.11} in the scalar case given in \cite{bt} is based in the classical results of the scalar Gel'fand--Levitan method \cite{gl},\cite{lev}, \cite{lg}, \cite{mar}, and among other results, in Lemma 2  of \cite{bt}. For the proof of Lemma 2 of \cite{bt} it is claimed that the difference between a   potential  and the potential obtained after removing one eigenvalue  is monotonic for large $x.$ See however, the comments in page 55 of \cite{schimm} concerning  the validity of the monotonicity claimed in the proof of Lemma 2 of \cite{bt}. In our proof of Theorem~\ref{theorem1} we proceed in a different way. We first prove that the proof of Theorem~\ref{theorem1} can be reduced to the proof in the particular case of potentials of compact support. Then, we prove Theorem~\ref{theorem1} for potentials of compact support using our results in transformations to remove eigenvalues  of matrix Schr\"odinger operators on the half-line \cite{AW2024}. The paper is organized as follows. In Section~\ref{section2} we state results from \cite{AW2021} on the matrix Schr\"odinger operator on the half-line that we use. In Section~\ref{section4} we state the results from \cite{AW2024}  in transformations to remove  eigenvalues that we need. Finally, in Section~\ref{inverse} we prove Theorem~\ref{theorem1}.
  
  The matrix Schr\"odinger equations have been studied since the early days of quantum mechanics. They are essential to consider properties of particles, such as spin, as well as to consider collections of particles. They have applications, for example, in nuclear, atomic,  and molecular physics. An important example is the Pauli equation, that is the Schr\"odinger equation of a spin one half particle. For these applications see, for example, the monographs  \cite{CS1989} and \cite{ll}.  The theory of quantum graphs gave a new impetus to the interest in matrix Schr\"odinger equations.  Quantum graphs  have important applications in several areas, including  nanotechnology,  quantum wires, and  quantum computing.  A star graph, that is to say a quantum graph with only one vertex, and a finite number of semi-infinite edges that meet at the vertex, is the particular case of a matrix Schr\"odinger equation where the potential $V$ is a diagonal matrix. For  a general introduction to quantum graphs, as well as for many results and applications, the readers can consult  the monographs  \cite{bk}, \cite{ku}. In the monograph \cite{AW2021} the readers can find more information  about  applications of matrix Schr\"odinger equations, as well as on  the literature.

\section{The half-line matrix Schr\"odinger equation}
\label{section2}

In this section we introduce  preliminary results that we need later. In \cite{AW2021} the readers can find 
 further information on the half-line matrix Schr\"odinger equation.
Consider the half-line matrix Schr\"odinger equation,
\begin{equation} 
\label{2.1}
-\psi(x)''+V(x)\,\psi(x)=k^2\psi(x),\qquad  k \in \mathbb C,  x\in\bR^+,
\end{equation}
where the prime denotes the $x$-derivative, the
potential $V(x)$ is an $n\times n$ selfadjoint matrix-valued function
of $x.$
The wavefunction
$\psi(x)$ is either an $n\times n$ matrix or
a column vector with
$n$ components.
We denote
$\bR^+:=(0, \infty).$  The selfadjointness of the potential means that,
\begin{equation}
\label{2.2}
V(x)^\dagger=V(x),\qquad x\in\bR^+.
\end{equation}
By  the dagger we denote the matrix adjoint.
We always assume that the potential $V$ is integrable, i.e. it satisfies \eqref{1.4b}.
 In some cases we suppose that the potential $V$ belongs to the Faddeev class
$L^1_1(\mathbb R^+).$
Namely, that,
\beq\label{2.4}
\int_0^\infty \,dx \,  (1+x)  |V(x)| < \infty.
\ene
 We use $\bCp$ to denote the upper half of the complex plane $\mathbb C,$ and use
$\bR$ for the real axis. We let $\bCpb:=\bCp\cup\bR.$

As we alredy mentioned in the introduction,  we denote by $H_{A,B}(V)$ the selfadjoint realization in $L^2(\mathbb R^+, \mathbb C^n)$ of the formal Schr\"odinger operator $-\frac{d^2}{dx^2}+ V(x)$ with the boundary condition \eqref{1.5}, where the boundary matrices $A,B$ satisfy \eqref{1.6}, \eqref{1.7}.

Of particular importance are the following two matrix solutions to \eqref{2.1}. The first one is  the Jost solution
$f(k,x)$ that satisfies the
asymptotic condition
\begin{equation}
\label{2.9}
f(k,x)=e^{ikx}\left[ I+o(1)\right], f'(k,x)= e^{ikx} \left[ik I+o(1)\right],
\qquad x\to \infty,
\end{equation}
for $ k \in \overline{\mathbb C^+}\setminus\{0\}.$
We denote by  $I$ the $n\times n$
identity matrix.  
Further, if  $ V\in L^1_1(\mathbb R^+)$ the Jost solution exists also at $k=0.$
The  second  important   matrix  solution to \eqref{2.1}  is the regular solution $\varphi(k,x), k \in \mathbb C$ that satisfies 
the initial conditions
\begin{equation}
\label{2.10}
\varphi(k,0)=A,\quad \varphi'(k,0)=B.
\end{equation}
Recall that  $A$ and $B$ are the boundary matrices appearing in
\eqref{1.5}.
Note  that $f(k,x)$ does not satisfy, in general,
the boundary condition \eqref{1.5}.  However, the regular solution
$\varphi(k,x)$ does  satisfy \eqref{1.5}.

We define the Jost matrix associated with \eqref{1.5} 
\eqref{2.1}   as follows.
\begin{equation}
\label{2.11}
J(k):=f(-k^*,0)^\dagger\,B-f'(-k^*,0)^\dagger\,A,\qquad k\in\mathbb R\setminus \{0\}.
\end{equation}
The asterisk denotes complex conjugation. If $V \in L^1_1(\mathbb R^+)$ the Jost matrix can be defined also at $k=0.$ The Jost matrix is an
$n\times n$ matrix-valued function of
$k.$ Moreover, it has an extension  to $\bCp,$ where the asterisk in
\eqref{2.11} is used to indicate how that
extension occurs.

We discuss now the  bound states of the half-line matrix Schr\"odinger operator.
For a given $k$ a bound-state solution corresponds to a square integrable, column-vector solution
to \eqref{2.1} that satisfies the boundary condition \eqref{1.5}. We  denote $\lambda:=k^2,  k \in \overline{\mathbb C^+}.$   
 The real number $\lambda=k^2$ is an eigenvalue of $H_{A,B}(V)$ if an only if  for the corresponding $k$  \eqref{2.1} has a bound-state solution.  By Theorem~3.11.1 of \cite{AW2021}, there are no bound states when $\lambda >0$ but  it is possible that there is a bound state at $\lambda=0.$   Further, if $V \in L^1_1(\mathbb R^+)$ by
Theorem~3.11.1 of \cite{AW2021}  for $\lambda=0$  there is no bound state, and
  the number of negative bound states is finite. Moreover, the multiplicity of the bound states is smaller or equal to $n.$   The  bound states when $\lambda<0$ appear at the $k$-values on the positive imaginary axis  of the complex $k$-plane that correspond to the zeros of $\det[J(k)].$
We use $\det[J(k)]$ to designate the determinant of the
Jost matrix $J(k).$ 
We suppose that there are
$N$ zeros of $\det[J(k)]$ that appear when $k=i\kappa_j$ for
$j=1, \dots,$ with
$\kappa_j$ being distinct positive
numbers. Remark that $N$ is equal to  the number of negative bound states
without counting multiplicities. The quantity $N$ can be zero,  a positive number or $\infty.$  
Hence, $\det[J(i\kappa_j)]=0$ and we denote by
$m_j$  the dimension of  $\text{\rm{Ker}}[J(i\kappa_j)].$ The quantity $m_j$ coincides with the multiplicity of the bound state at $ k=i \kappa_j.$

Following \cite{AW2024}, we use the Gel'fand--Levitan theory  to analyze  the  bound-state solutions to
\eqref{2.1} 
In this theory  the normalization matrices for the bound states are obtained by
normalizing the regular solution $\varphi(k,x)$ at the bound states.
 We denote by  $C_j$ and $\Phi_j(x)$  the
Gel'fand--Levitan normalization matrix and
the corresponding normalized matrix solution
at the bound state with $k=i\kappa_j,$ respectively.
We proceed to define the $n\times n$ matrices $C_j$ and $\Phi_j(x)$ following \cite{AW2024}.
Let us use $Q_j$ to denote the orthogonal projection
onto $\text{\rm{Ker}}[J(i\kappa_j)].$ 
 The Gel'fand--Levitan bound-state normalized solution
 to the Schr\"odinger equation is defined as follows,
\begin{equation}
\label{3.28b}
\Phi_j(x):=\varphi(i\kappa_j,x)\,C_j,
\end{equation}
where the Gel'fand--Levitan  normalization matrix $C_j$  is defined below.
Each of $C_j$  is a nonnegative matrix of rank $m_j,$ such that $\Phi_j(x)$ is square-integrable, and moreover,
\begin{equation}
\label{3.29}
\Phi_j(x)=O(e^{-\kappa_j x}),\qquad x\to \infty.
\end{equation}
By \eqref{2.10}, $\Phi_j(x)$ satisfies \eqref{1.5}. 
 We also have that  the following normalization conditions hold,
\begin{equation}
\label{3.30}
\int_0^\infty dx\,\Phi_j(x)^\dagger\,\Phi_l(x)=Q_j  \delta_{j,l}  ,\qquad  j,l=1,\dots,
\end{equation}
where we denote by $\delta_{j,l}$ the Kronecker delta. To construct the $n\times n$ Gel'fand--Levitan normalization matrix $C_j$ we define
 the $n\times n$ matrix $\mathbf G_j$ as
\begin{equation}
\label{3.32}\mathbf G_j:=\int_0^\infty dx\,Q_j\,\varphi(i\kappa_j,x)^\dagger\,\varphi(i\kappa_j,x)\,Q_j.
\end{equation}
By Theorem 3.11.1 (e) of \cite{AW2021}  the integral in the right-hand side of \eqref{3.32} is finite.
Further, we introduce the matrix $\mathbf H_j$ as
\begin{equation}
\label{3.33}
\mathbf H_j:=I-Q_j+\mathbf G_j.
\end{equation}
Both $\mathbf G_j$ and $\mathbf H_j$ are 
selfadjoint. Moreover, $\mathbf H_j$ is a positive matrix, and we denote by $\mathbf H_j^{1/2}$ its unique positive square root. Furthermore $\mathbf H_j$ sends $ Q_j \mathbb C^n$ into   $ Q_j \mathbb C^n$    and the restriction of 
$\mathbf H_j$ to  $ Q_j \mathbb C^n$ is also positive.
The Gel'fand--Levitan normalization matrix $C_j$ is defined
as
\begin{equation}
\label{3.35}
C_j:=\mathbf H_j^{-1/2}\,Q_j,\qquad j=1,\dots.
\end{equation}
 We have  that $C_j$ is selfadjoint and nonnegative, and it has rank
 equal to $m_j,$ the same as the rank of $Q_j.$ Moreover,
 \beq\label{3.35b}
 Q_j C_j= C_j Q_j= C_j, \qquad j=1,\dots.
 \ene

 \section{  Transformation to remove a bound state}
\label{section4}
In this section we state  results from  Section 6 of \cite{AW2024} in a transformation to remove a bound state. We state the results for integrable potentials with compact support, that is the case that we use in this paper. For more general results see Section 6 of \cite{AW2024}. 
We remove any one of the bound states with $\lambda=\lambda_j,$ where $\lambda_j:=-\kappa_j^2,$ and with
the Gel'fand--Levitan normalization matrix $C_j.$
Note that as we  do not order the distinct positive constants $\kappa_j$ in any particular way, without loss
of generality we can suppose that we remove the bound state
with $k=i\kappa_N$ and the Gel'fand--Levitan normalization matrix  $C_N.$
After that, we obtain
the perturbed Schr\"odinger operator with 
the potential
$\tilde V(x),$ the boundary matrices $\tilde A$ and $\tilde B,$ the regular solution
$\tilde\varphi(k,x),$ the Jost matrix $\tilde J(k),$  and
$N-1$ bound states with eigenvalues $-\tilde\kappa_j^2,$ 
the Gel'fand--Levitan normalization matrices $\tilde C_j,$ the orthogonal projections
$\tilde Q_j$ onto $\text{\rm{Ker}}[\tilde J(i\kappa_j)],$ 
the Gel'fand--Levitan normalized bound-state solutions $\tilde\Phi_j(x),$ and the multiplicities $\tilde m_j$ of the bound states, for $j=1,\dots, N-1.$
In the next theorem, that summarizes results from  Section 6 of \cite{AW2024} in the case of potentials with compact support, we express the perturbed quantities distinguished with a tilde
in terms of the unperturbed quantities not containing the tilde and the perturbation identified with $\kappa_N$ and $C_N.$ However, before we state the theorem we introduce the Moore--Penrose inverse.

We designate by $M^+$  the Moore--Penrose inverse
of a matrix $M,$ \cite{BG2003,CM2009}.
We only deal with Moore--Penrose inverses of square matrices.
As stated in Definitions~1.12 and 1.13 and Theorem~1.1.1 of \cite{CM2009},
the matrix $M^+$ is the Moore--Penrose inverse of the matrix $M$ if  the following four equalities are fulfilled:
\begin{equation}
\label{3.3}
\begin{cases} M M^+ M=M,\quad M^+ M M^+=M^+,\\
\noalign{\medskip}
 (M M^+)^\dagger=M M^+,\quad
(M^+ M)^\dagger=M^+ M.
\end{cases}
\end{equation}

\begin{theorem}\label{theorem6.1}
Consider
the unperturbed Schr\"odinger operator with the potential
$V$ satisfying \eqref{1.4b}, \eqref{2.2},  and with support in the  interval  $[0,x_0]$. Further,  the selfadjoint
boundary condition \eqref{1.5} is described by the boundary matrices $A$ and $B$ satisfying
\eqref{1.6} and \eqref{1.7}, 
with the regular solution $\varphi(k,x)$ satisfying the initial conditions
 \eqref{2.10}, the Jost solution $f(k,x)$ satisfying \eqref{2.9},
 the Jost matrix $J(k)$ defined in \eqref{2.11}, 
 containing
$N\geq 1$ bound states with  eigenvalues  $\lambda_j=-\kappa_j^2,$ 
the Gel'fand--Levitan normalization matrices $C_j,$ 
the orthogonal projections
$Q_j$ onto $\text{\rm{Ker}}[J(i\kappa_j)],$ 
and
the Gel'fand--Levitan normalized bound-state solutions $\Phi_j(x)$ for $1\le j\le N.$ 

Let us denote by  $W_N(x),$ 
\begin{equation}
\label{6.3}
W_N(x):=\int_x^\infty dz\, \Phi_N(z)^\dagger\,\Phi_N(z),
\end{equation}
and define
the matrix-valued perturbed potential $\tilde V(x)$ as
\begin{equation}
\label{6.24}
\tilde V(x):=V(x)+2\,\ds\frac{d}{dx}\left[
\Phi_N(x) \,W_N(x)^+\,\Phi_N(x)^\dagger
\right],\end{equation} 
where we recall that $W_N(x)^+$ denotes the Moore--Penrose inverse of $W_N(x).$ Then, we have:
\begin{enumerate}
\item[\text{\rm(a)}] The perturbed potential $\tilde V(x)$ appearing in \eqref{6.24} 
satisfies  \eqref{1.4b} and\eqref{2.2}. Moreover, its support is contained in the interval $[0,x_0].$

\item[\text{\rm(b)}] 
The quantity, 
$$
\varphi(k,x)= \tilde{\varphi}(k,x)+ \Phi_N(x)    W_N(x)^+\int_0^x \, dy\,   \Phi_N(y)^\dagger     \tilde{\varphi}(k,y) dy,
$$
is a solution to \eqref{2.1} with the potential \eqref{6.24}.

\item[\text{\rm(c)}] For $ k \neq \pm  i \kappa_N,$  the perturbed quantity $\tilde\varphi(k,x)$ can be expressed as
\begin{equation}
\label{6.44}
\tilde\varphi(k,x)=\varphi(k,x)+\ds\frac{1}{k^2+\kappa_N^2}
\,\Phi_N(x) \,W_N(x)^+\left[\Phi'_N(x)^\dagger\,\varphi(k,x)-\Phi_N(x)^\dagger
\,\varphi'(k,x)\right].
\end{equation}

\item[\text{\rm(d)}] Under the perturbation,
the projection matrices $Q_j$ for $1\le j\le N-1$ remain unchanged, i.e. we have
 \begin{equation}
\label{6.114}
\tilde Q_j=Q_j,\qquad 1\le j\le N-1.\end{equation}

\item[\text{\rm(e)}] Under the perturbation, the Gel'fand--Levitan normalization matrices  for $1\le j\le N-1$ remain
 unchanged, i.e. we have
 \begin{equation}
\label{6.115}
\tilde C_j=C_j,\qquad 1\le j\le N-1.\end{equation}

\item[\text{\rm(f)}] The perturbed quantity $\tilde\varphi(k,x)$ satisfies the initial conditions \eqref{2.10} with $A$, $B$  replaced by $\tilde A,$  $\tilde B,$ respectively,  and
where the matrices $\tilde A$ and $\tilde B$
are expressed in terms of the unperturbed
boundary matrices $A$ and $B$ and
the Gel'fand--Levitan normalization matrix $C_N$ for the bound state at $k=i \kappa_N$ as
\begin{equation}\label{6.45}
\tilde A=A, \quad \tilde B=B+A\,C_N^2 A^\dagger A.
\end{equation}

\item[\text{\rm(g)}] 
The matrices $\tilde A$ and $\tilde B$ appearing in \eqref{6.45} satisfy \eqref{1.6} and \eqref{1.7}. Hence, 
as a consequence of (b) and (f), 
the quantity $\tilde\varphi(k,x)$ is the regular solution 
to the matrix Schr\"odinger equation with the potential $\tilde V(x)$ in \eqref{6.24} 
and with the selfadjoint
boundary condition \eqref{1.5} with $A$ and $B$  replaced with
$\tilde A$ and $\tilde B,$ respectively.

\item[\text{\rm(h)}] Under the perturbation, the determinant of the Jost matrix is transformed as
 \begin{equation}
\label{6.82}
\det[\tilde J(k)]=\left(\ds\frac{k+i\kappa_N}{k-i\kappa_N}\right)^{m_N} \det[J(k)],\qquad
k\in\overline{\mathbb C^+},\end{equation}
where we recall
that $m_N$ is the multiplicity of the bound state of the unperturbed problem at $k=i\kappa_N.$

\item[\text{\rm(i)}] Under the perturbation,
the bound state with eigenvalue $\lambda_N=-\kappa_N^2$ is removed without adding any new bound states in such a way that
the remaining bound states with eigenvalues $\lambda_j=-\kappa_j^2$ and their multiplicities $m_j$ for $1\le j\le N-1$
are unchanged.

\item[\text{\rm(j)}] Under the perturbation the absolutely continuous spectrum remains unchanged and equal to $[0,\infty).$ Moreover, the spectral measures for the absolutely continuous spectrum of the unperturbed and the perturbed problems are the same. For the definition of the spectral measure see \cite{AW2024}
\end{enumerate}
\end{theorem}
\section{Reverse Lieb--Thirring inequality}
\label{inverse}
In this section we give the proof of Theorem~\ref{theorem1}.
We first prove that we can reduce the problem to the case of potentials with compact support.  Recall that by Theorem 3.11.1 (g) of \cite{AW2021} for potentials in $L^1_1(\mathbb R^+),$ and in particular for potentials of compact support,  the number of negative eigenvalues is finite. Let the potential $V$ belong to $L^1(\mathbb R^+)$ and let us denote by $V_\pm: 1/2(|V|\pm V)$,  its positive, respectively, negative part. Hence,
\beq\label{3.1}
V(x)= V_+(x)- V_-(x), \qquad x \in \mathbb R^+.
\ene
We denote by $\chi_{[0,l]}(x), l=1,\dots,$ the characteristic function of $[0,l],$ and we define,
\beq\label{3.2}
V_l(x) = V_+(x)-   \chi_{[0,l]}(x) V_-(x).
\ene
Recall that  $H_{I,B}(V)$ is the selfadjoint realization  of the matrix  Schr\"odinger operator $- \frac{d^2}{dx^2}+V(x)$ with the boundary condition \eqref{1.5} where the boundary matrices $I,B$ satisfy \eqref{1.6}, \eqref{1.7}. Recall that  this just amounts to ask that $B$ is selfadjoint. Similarly, we denote by $H_{I,B}(V_l)$ the Schr\"odinger operator with the potential $V$ replaced by $V_l.$  Let   $\lambda_j:= -\kappa_j^2,$ for $ j=1,\dots$ be the negative eigenvalues of $H_{I,B}(V),$ in increasing  order. Recall that  $m_j$ is the multiplicity of $\lambda_j.$ Similarly, we denote by   $\lambda_j^{(l)}:= -\kappa_{j,l}^2,$ for $ j=1,\dots,  l=1,\dots,$  the negative eigenvalues of $H_{I,B}(V_l),$ also in increasing order , and by $m_j^{(l)}$ 
the multiplicity of   $\lambda_j^{(l)}.$ Let $\mu_j,$ for $j=1,\dots, $ denote the negative eigenvalues of $H_{I,B}(V),$  in nondecreasing order and repeated according to its  multiplicity. Further, let  $\mu_j^{(l)},$ for  $j=1,\dots,$ designate the negative eigenvalues of $H_{I,B}(V_l)$ in nondecreasing order and repeated according to its multiplicity. Since $ H_{I,B}(V)\leq  H_{I,B}(V_l) ,$ it follows from the min-max principle \cite{rs4} that,
\beq\label{3.3b}
\mu_j \leq \mu_j^{(l)}, \qquad j=1,\dots.
\ene
Then, by \eqref{3.3b}
\beq\label{3.4}
\sum_{j}   m_j  \ds \sqrt{\left|\lambda_j\right|} \geq  \sum_{j}  m_j^{(l)} \sqrt{\left|\lambda_j^{(l)}\right|} , \qquad l=1,\dots.
\ene 
 Below  we use \eqref{3.4} to reduce the proof of the reverse Lieb--Thirring inequality for $H_{I,B}(V)$  to the proof of the reverse Lieb--Thirring inequality for $H_{I,B}(V_l).$ 

We first   prove that $H_{I,B}(V_l)$ has a finite number of  negative eigenvalues. Let us denote,
\beq\label{3.4b}
V_{l, p}:= \chi_{[0,p]}(x) V_l(x), \qquad l,p= 1,\dots,
\ene
and by $H_{I,B}(V_{l,p})$ the Schr\"odinger operator defined as  $H_{I,B}(V)$ with the potential $V_{l,p}$ instead of $V.$ As  $V_{l,p}$ has compact support the operator $H_{I,B}(V_{l,p})$ has a finite number of negative eigenvalues.  We designate by $\mu_j^{(l,p)},$ for $ j=1,\dots,$ the negative eigenvalues of   $H_{I,B}(V_{l,p})$ in nondecreasing order and repeated according to its multiplicity. Since for
$p \geq l, H_{I,B}(V_{l,p}) \leq   H_{I,B}(V_{l}),$ by the min-max principle \cite{rs4}
\beq\label{3.4c}
\mu_j^{(l,p)} \leq \mu_j^{(l)}, \qquad p \geq l, j=1,\dots.
\ene
Assume that for some fixed $l=1,\dots, H_{I,B}(V_{l})$ has an infinite number of negative eigenvalues. Then, by min-max principle \cite{rs4} there would be an infinite number of negative $\mu_j^{(l)}.$ But by \eqref{3.4c} this would imply the existence of an infinite number of negative  $\mu_j^{(l,p)} $ for all $p \geq l.$ However, this is impossible, again by the the
min-max principle \cite{rs4}, because as $V_{l,p}$ has compact support, $H_{I,B}(V_{l,p})$ has only a finite number of negative eigenvalues. It follows that $H_{I,B}(V_{l})$ has a finite number of negative eigenvalues. 

We now prove that for $l$ large enough  $H_{I,B}(V_l)$ has at least one negative eigenvalue, using that  $H_{I,B}(V)$ has at least one negative eigenvalue. For this purpose, we prove that $H_{I,B}(V_l)$ converges to $H_{I,B}(V)$ in norm resolvent sense. Let us introduce the polar decomposition for  $V(x),$
\beq\label{3.5}
V(x)= U(x) \hat{V}(x),
\ene
where $U(x)$ is a partially isometric matrix and $\hat{V}(x)$  is the absolute value of $V(x),$ a selfadjoint nonnegative matrix. For a concrete representation of $U(x)$ and of  $ \hat{V}(x)$ see equation (4.2.24) to (4.2.26) of \cite{AW2021}. Further, denote
\beq\label{3.6}
V^{(1)}:=  \sqrt{\hat{V}(x)}, \qquad  V^{(2)}:=  U(x)\, \sqrt{\hat{V}(x)}.
\ene
We designate by
\beq\label{3.7}
R_{0,I,B}(z):= \left(H_{I,B}(0)-z\right)^{-1},\qquad z \in \rho(H_{I,B}(0)),
\ene
the resolvent of $H_{I,B}(0),$ where $\rho(H_{I,B}(0))$ denotes the resolvent set of $H_{I,B}(0).$ Further, we introduce the resolvent of $H_{I,B}(V),$
\beq\label{3.8}
R_{I,B,V}(z):= \left(H_{I,B}(V)-z\right)^{-1},\qquad z \in \rho(H_{I,B}(V)),
\ene
where $\rho(H_{I,B}(V))$ denotes the resolvent set of  $H_{I,B}(V).$ Then, by equation (4.2.39) of \cite{AW2021},
\beq\label{3.9}
R_{I,B,V}(z)= R_{0,I,B}(z)-R_{0,I,B}(z)\left(I+ V^{(1)}  R_{0,I,B}(z) V^{(2)}\right)^{-1}R_{0,I,B}(z),
\ene
for $z \in \rho(H_{I,B}(0))\cap \rho(H_{I,B}(V)).$ In a similar way, replacing in the formulae above $V$ by $V_l,$  and $H_{I,B}(V)$ by $H_{I,B}(V_l)$  we get,
\beq\label{3.10}
R_{I,B,V_l}(z)= R_{0,I,B}(z)-R_{0,I,B}(z)\left(I+ V^{(1)}_l  R_{0,I,B}(z) V^{(2)}_l\right)^{-1}R_{0,I,B}(z),
\ene
for $z \in \rho(H_{I,B}(0))\cap \rho(H_{I,B}(V_l)).$ By equation (4.2.11) of \cite{AW2021}, and as $V\in L^1(\mathbb R^+),$ the operators
$$
 V^{(1)}  R_{0,I,B}(z) V^{(2)}, \qquad  V^{(1)}_l  R_{0,I,B}(z) V^{(2)}_l,
$$
are Hilbert Schmidt, and
\beq\label{3.11}
\lim_{l\to \infty}   V^{(1)}_l  R_{0,I,B}(z) V^{(2)}_l=  V^{(1)}  R_{0,I,B}(z) V^{(2)},
\ene
where the limit is in the Hilbert-Schmidt norm. Then, by \eqref{3.9},\eqref{3.10}, and \eqref{3.11} $H_{I,B}(V_l)$ converges in norm resolvent sense to $H_{I,B}(V).$   The smallest negative eigenvalue of  $H_{I,B}(V)$ is separated from the rest of the spectrum  by a small circle  $\Gamma.$  Then, it follows from Theorems 2.25 and 3.16, and the comments in Section 5, of Chapter IV, of \cite{kato},   that for $l$ large enough $H_{I,B}(V_l)$ has at least one negative eigenvalue inside the circle $\Gamma.$

We now prove that we can reduce the proof of the reverse  Lieb--Thirring inequality for   $H_{I,B}(V_l)$ to the proof of the  reverse Lieb--Thirring inequality for   $H_{I,B}(V_{l,p}).$  For this purpose we observe that for each fixed $l$ 
 the operator   $H_{I,B}(V_{l,p})$ converges to   $H_{I,B}(V_{l})$ as $ p \to \infty$ in norm resolvent sense. The proof is the same as the proof that    $H_{I,B}(V_{l})$ converges to   $H_{I,B}(V)$ as $ l \to \infty$ in norm resolvent sense that we gave above. Take any $ \varepsilon >0$ small enough so that each negative eigenvalue $\lambda_j^{(l)}$   is separated by an open  disk of center $\lambda_j^{(l)}$ and  radius $\varepsilon$ from the rest of the spectrum. Note that this is possible because    $H_{I,B}(V_l)$ has a finite number of negative eigenvalues.  Hence, using again 
  Theorems 2.25 and 3.16, and the comments in Section 5, of Chapter IV, of \cite{kato}, we get that  for $ p$ large enough, inside the disk  of center $\lambda_j^{(l)}$ and  radius $\varepsilon$ the operator  $H_{I,B}(V_{l,p})$ has a finite number of negative eigenvalues of total multiplicity $m_j^{(l)}$ and  outside the union of these open disks  $H_{I,B}(V_{l,p})$ has no negative eigenvalues. Since $\varepsilon$ can be chosen arbitrarily small we have,
 \beq\label{3.12}
 \lim_{p\to \infty} \sum_{j}  m_j^{(l,p)}  \lambda_j^{(l,p)}= \sum_{j}  m_j^{(l)}  \lambda_j^{(l)},
\ene
where we denote by$\lambda _j^{(l,p)},$ for $ j=1,\dots$ the negative eigenvalues of  $H_{I,B}(V_{l,p}),$  in increasing order, and by $m _j^{(l,p)},$ for  $j=1,\dots$ the multiplicity of the negative eigenvalue  $\lambda _j^{(l,p)}.$  
Below we prove the reverse Lieb--Thirring inequality for $H_{A,B}(V_{l,p}),$ and we use \eqref{3.12}
to obtain the reverse Lieb--Thirring inequality for  $H_{I,B}(V_{l}).$  

Removing the  negative eigenvalues of  $H_{I,B}(V_{l,p}),$ one by one, as in Theorem~\ref{theorem6.1}, we obtain the operator $H_{I, \tilde{B}}(\tilde{V}_{l,p})$ with no negative eigenvalues, where
\beq\label{3.14}
\tilde{V}_{l,p}(x)= V_{l,p}(x)+ 2 \sum_{j}  \,\ds\frac{d}{dx}\left[
\Phi_j^{(l,p)}(x) \,\left(W_j^{(l,p)}(x)\right)^+\,\left(\Phi_j^{(l,p)}(x)\right)^\dagger\right],
\ene
where $\Phi_j^{(l,p)}(x)$ is the Gel-fand--Levitan normalized matrix solution for the eigenvalue $\lambda_j^{(l,p)}= - \left(\kappa_j^{(l,p)}\right)^2$ of    $H_{I,B}(V_{l,p}),$ and $W_j^{(l,p)}(x)$ is defined as in \eqref{6.3}, but with $\Phi_j^{(l,p)}(x)$ instead of $\Phi_N(x).$ Further,
\beq\label{3.15}
 \tilde{B}=B+\sum_{j}\left[C_j^{(l,p)}\right]^2,
 \ene
where $C_j^{(l,p)}$ is the Gel-fand--Levitan normalization matrix for the negative eigenvalue $\lambda_j^{(l,p)}$ of    $H_{I,B}(V_{l,p}).$  Note that by \eqref{3.30}
\beq\label{3.16}
W_j^{(l,p)}(0)=  {Q}_j^{(l,p)},\qquad j=1, \dots,
\ene
with  ${Q}_j^{(l,p)}$  the orthogonal projection onto the kernel of ${J}^{(l,p)}\left(i \kappa_j^{(l,p)}\right).$ The quantity 
 ${J}^{(l,p)}(k)$ is the Jost matrix of   $H_{I,B}(V_{l,p}).$  Further, we used that  $\left({Q}_j^{(l,p)}\right)^+= {Q}_j^{(l,p)}$ as it can be easily verified using the definition \eqref{3.3} and that  ${Q}_j^{(l,p)}$ is an orthogonal projection.   Moreover, by equations (6.98), (6.99), (6.108), and (6.110) of
 \cite{AW2024}
 \beq\label{3.17}
 \lim_{x\to\infty} \left[
\Phi_j^{(l,p)}(x) \,\left(W_j^{(l,p)}(x)\right)^+\,\Phi_j^{(l,p)}(x)^\dagger\right]= 2 \kappa_j^{(l,p)} P_j^{(l,p)}, \qquad  j=1, \dots,
\ene
where $ P_j^{(l,p)}$ is the orthogonal projection onto the kernel of  $\left({J}^{(l,p)}\left(i \kappa_j^{(l,p)}\right)\right)^\dagger.$  The dimension of $\text{\rm{Ker}}\left[\left(J^{(l,p)}\left(i\kappa_j^{(l,p)}\right)\right)^\dagger\right]$ is $m_j^{(l,p)},$
that is also the dimension of $\text{\rm{Ker}}\left[J^{(l,p)}\left(i\kappa_j^{(l,p)}\right)\right].$

 By \eqref{3.35b}, \eqref{3.14}, \eqref{3.16}, and \eqref{3.17},
 \beq\label{3.18}
\int_0^\infty\, \tilde{V}_{l,p}(x)\, dx  = \int_0^\infty V_{l,p}(x)\, dx - 2 \sum_{j}  \left[C_j^{(l,p)}\right]^2   +4
\sum_{j}\sqrt{\left|\lambda_j^{(l,p)}\right|} \ds \, P_j^{(l,p)}.
\ene
For later use we  prove that
\beq\label{3.19}
\int_0^\infty \text{\rm Tr}[ \tilde{V}_{l,p}(x)]\, dx+\text{\rm Tr}[ \tilde{B}]\geq 0.
\ene
This statement was proved in the scalar case in \cite{bt} using results in the  Gel'fand-Levitan method that  are known in the scalar case. Here we prove that \eqref{3.19} is an immediate consequence of the fact that as $H_{I, \tilde{B}}(\tilde{V}_{l,p})$ has no negative eigenvalues it is a nonnegative operator. Let us denote by $H^1(\mathbb R^+, \mathbb C^n)$ the  Sobolev space of all functions in $L^2(\mathbb R^+, \mathbb C^n),$ with the first derivative in $L^2(\mathbb R^+,\mathbb C^n).$ Then, the quadratic form of $H_{I, \tilde{B}}(\tilde{V}_{l,p})$ is given by,
\beq\label{3.20}
q_{I, \tilde{B}, \tilde{V}_{l,p}}(\phi, \psi):= \sum_{i=1}^n \left(\phi_j', \psi_j '\right)+(\tilde{V}_{l,p}\phi,\psi) +\phi^\dagger(0) \tilde{B}\psi(0),\qquad
\phi, \psi \in H^1(\mathbb R^+, \mathbb C^n),
\ene
with domain $H^1(\mathbb R^+, \mathbb C^n).$ Let $ f\in C^1([0,\infty))$ be real valued and satisfy $f(x)=1, 0 \leq x \leq 1, f(x)=0, x \geq 2.$ For $r=1,\dots,n, s=1,\dots,$ denote,
$$
\phi^{(r,s)}:= (0,\dots, f(x/s),0,\dots, 0)^T,
$$  
with  $ f(x/s)$ in the $r$ position. Moreover, since $H_{I, \tilde{B}}(\tilde{V}_{l,p})$ has no negative eigenvalues the quadratic form \eqref{3.20} is nonnegative, and hence,
\beq\label{3.21}
q_{I, \tilde{B}, \tilde{V}_{l,p}}(\phi^{(r,s)},\phi^{(r,s)} ):=  \frac{1}{s} \int_0^\infty (f'(y))^2\,dy +\int_0^\infty \left( \tilde{V}_{l,p}\right)_{r,r}(x)\, f^2(x/s) \,dx+
 \tilde{B}_{r,r}\geq 0, 
\ene
where we denote by $\left( \tilde{V}_{l,p}\right)_{r,r}(x)$ the $r,r$ entry of $ \tilde{V}_{l,p}(x),$ and by $\tilde{B}_{r,r}$ the $r,r$ entry of $B.$
Taking the limit as $ s \to \infty$ in \eqref{3.21} we get,
\beq\label{3.22}
\int_0^\infty  \,dx\, \left( \tilde{V}_{l,p}\right)_{r,r}(x) +
 \tilde{B}_{r,r}\geq 0, \qquad r=1,\dots,n. 
 \ene
 Equation \eqref{3.19} follows from \eqref{3.22}. Furthermore, by \eqref{3.15}, \eqref{3.18} and \eqref{3.19},
 \beq\label{3.23}
  \sum_{j}   m_j^{(l,p)} \sqrt{\left|\lambda_j^{(l,p)}\right|} \ds  \geq \frac{1}{4}\left [
 - \int_0^\infty \text{\rm Tr}\left[V_{l,p}(x)\right]\, dx - \text{\rm Tr}[B]+  \sum_{j}\text{\rm Tr}\left[ \left(C_j^{(l,p)}\right)^2 \right] \right].
\ene
We now prove that the last term in the right-hand side of \eqref{3.23} is bounded below  by a positive constant uniformly in $l$ and $p.$  For this purpose, it is enough to prove that this is so for $\text{\rm Tr}\left[ \left(C_1^{(l,p)}\right)^2 \right].$ Let $v_j^{(l,p)},$ for $ j=1,\dots,m_1^{(l,p)}$ be an orthonormal basis of the kernel of $ J^{l,p}\left(i \sqrt{\left|\lambda_1^{(l,p)}\right|}\right).$ Then,
\beq \label{3.24}
Q_1^{(l,p)}= \sum_{j=1}^{m_1^{(l,p)}}  v_j^{(l,p)} \left(v_j^{(l,p)}\right)^\dagger.
\ene
Let us denote by $\varphi^{(l,p)}(k,x),$ respectively, $f^{(l,p)}(k,x)$,  the regular solution and the Jost solution for the potential $V_{l,p}$ with the boundary condition \eqref{1.10}. By Theorem 3.11.1 of \cite{AW2021}
\beq\label{3.25}
\varphi^{(l.p)}\left(i  \sqrt{\left|\lambda_1^{(l,p)}\right|},x\right) Q_1^{(l,p)}= \sum_{j=1}^{m_1^{(l,p)}}  f^{(l.p)}\left(i  \sqrt{\left|\lambda_1^{(l,p)}\right|},x\right)
\omega^{(l,p)}_j  \left(v_j^{(l,p)}\right)^\dagger,
\ene
where $ w_j^{(l,p)}$ belongs to the kernel of  $ J^{l,p}\left(i \sqrt{\left|\lambda_1^{(l,p)}\right|}\right)^\dagger,$  and
\beq\label{3.26}
\varphi^{(l.p)}\left(i  \sqrt{\left|\lambda_1^{(l,p)}\right|},x\right) v_j^{(l,p)}= f^{(l.p)}\left(i  \sqrt{\left|\lambda_1^{(l,p)}\right|},x\right) w_j^{(l,p)}, \qquad j=1, \dots, m_1^{(l,p)}.
\ene
  It follows from the proof of Proposition 3.2.1 of \cite{AW2021} that the  Jost solution $f^{(l,p)}\left(i  \sqrt{\left|\lambda_1^{(l,p)}\right|},x\right)$ satisfies \eqref{2.9} with the $o(1)$ uniform in $l,p= 1,\dots.$ Further, it follows from the proof of Proposition 3.2.9 of \cite{AW2021} that the regular solution $\varphi^{(l.p)}\left(i  \sqrt{\left|\lambda_1^{(l,p)}\right|},x\right)$ is  bounded
in any interval $[0,x_0], x_0 >0,$ uniformly for all $l,p=1,\dots.$ Take an $x_0$  so large  that $f^{(l,p)}\left(i  \sqrt{\left|\lambda_1^{(l,p)}\right|},x_0\right)$ is invertible and $ |o(1)| < 1/2.$  Then,
\beq \label{3.27}
w_j^{(l,p)}= f^{(l,p)}\left(i  \sqrt{\left|\lambda_1^{(l,p)}\right|},x_0\right)^{-1} \varphi^{(l.p)}\left(i  \sqrt{\left|\lambda_1^{(l,p)}\right|},x\right) v_j^{(l,p)}.
\ene
It follows from \eqref{3.27},
\beq\label{3.28}
|w_j^{(l,p)}| \leq C, \qquad j=1,\dots m_1^{(l,p)},
\ene
where the constant $C$ is uniform in $l,p=1\dots.$ Then, by \eqref{2.9}, \eqref{3.32}, \eqref{3.25}, and \eqref{3.28},
\beq\label{3.29b}
|\mathbf G_1^{(l,p)}| \leq C,
\ene
where the constant $C$ is uniform in $l,p=1\dots.$ Further, by \eqref{3.33}, \eqref{3.35}, and \eqref{3.29b},
\beq\label{3.30b}
 \text{\rm Tr}\left[C_1^{(l,p)}\right]^2 \geq \delta >0,
\ene
where the positive constant $\delta$ is uniform on $l,p=1,\dots.$  By \eqref{3.23} and \eqref{3.30b}
\beq\label{3.31b}
  \sum_{j}   m_j^{(l,p)} \sqrt{\left|\lambda_j^{(l,p)}\right|} \ds  \geq \frac{1}{4}\left [
 - \int_0^\infty \text{\rm Tr}\left[V_{l,p}(x)\right]\, dx - \text{\rm Tr}[B]+  \delta  \right].
\ene
Moreover, by \eqref{3.12} and \eqref{3.31b}
\beq\label{3.32b}
  \sum_{j}   m_j^{(l)} \sqrt{\left|\lambda_j^{(l)}\right|} \ds  \geq \frac{1}{4}\left [
 - \int_0^\infty \text{\rm Tr}\left[V_{l}(x)\right]\, dx - \text{\rm Tr}[B]+  \delta  \right].
\ene
Finally by \eqref{3.4} and \eqref{3.32b},
\beq\label{3.33b}
 \sum_{j=1}^{N}   m_j^{} \sqrt{\left|\lambda_j\right|} \ds  > \frac{1}{4}\left [
 - \int_0^\infty \text{\rm Tr}\left[V_{l}(x)\right]\, dx - \text{\rm Tr}[B]  \right].
 \ene
 Let us prove that the constant $1/4$ in \eqref{1.11} is sharp. We consider the matrix Schr\"odinger operator
 $H_{I,0}(0)$ with the Neumann boundary condition $\psi'(0)=0,$ and potential identically zero. The operator 
  $H_{I,0}(0)$ has no eigenvalues.Then, using the results in Section 8 of \cite{AW2024} we add to  $H_{I,0}(0)$ a negative  eigenvalue, $\lambda_1= -\kappa_1^2, \kappa_1 >0,$ with multiplicity $m_1$ and with Gel'fand--Levitan norming constant $C_1,$ to obtain the matrix Schr\"odinger operator $H_{I,B}(V)$ with $ B=- C_1^2,$ and where $V$ is integrable and it satisfies $ V(x)=O(x e^{-2 \kappa_1 x}), x \to \infty.$ By equations  (8.17), (8.22), (8.37),(8.51), and (8.54) of \cite{AW2024}, since $ B=- C_1^2,$ and as we added a bound state to the identically zero potential,
   \beq\label{3.33c}
    \sqrt{\left|\lambda_1\right|} \ds \, P_1= \frac{1}{4} \left[-  \int_0^\infty V(x)\, dx - B +C_1^2 \right],
\ene
where we used that the orthogonal projection, $Q_1,$ onto the kernel of the Jost matrix, $J(k),$ of $H_{I,B}(V)$ at $ k =i\kappa_1,$ satisfies $Q_   1= Q_1^+.$ Recall that $P_1$ is the orthogonal projection onto the kernel of  $J(i\kappa_1)^\dagger.$   Taking traces in both sides of \eqref{3.33c} we obtain,

\beq\label{3.33d}
m_1 \sqrt{\left|\lambda_1\right|} \ds = \frac{1}{4} \left[-  \int_0^\infty\text{\rm Tr}[ V](x)\, dx -\text{\rm Tr}[B] +\text{\rm Tr}[C_1^2]\right].
\ene
Assume that \eqref{1.11} holds with $1/4$ replaced by $1/  \alpha,$ with 
$ 0< \alpha < 4,$ that is to say,
\beq\label{3.33e} 
  m_1  \sqrt{\left|\lambda_1\right|} \ds  > \frac{1}{\alpha}\left [
 - \int_0^\infty \text{\rm Tr}\left[V(x)\right]\, dx - \text{\rm Tr}[B]  \right].
 \ene
Then, by \eqref{3.33d} and \eqref{3.33e}
\beq\label{3.33f}
\left( \frac{1}{4}-\frac{1}{\alpha}\right) \left[-  \int_0^\infty\text{\rm Tr}[ V](x)\, dx -\text{\rm Tr}[B]\right]+ \frac{1}{4}
\text{\rm Tr}[C_1^2]  >0.
\ene
Introducing  \eqref{3.33d} in the left-hand side of \eqref{3.33f}  we get,
\beq\label{3.33g}
\left( \frac{1}{4}-\frac{1}{\alpha}\right) \left[4 m_1   \sqrt{\left|\lambda_1\right|} - \text{\rm Tr}\left[C_1^2\right]  \right]+
\frac{1}{4} \text{\rm Tr}\left[C_1^2]\right] >0.
\ene
Keeping $\lambda_1$ and $m_1$  fixed, and taking $\text{\rm Tr}[ C_1^2]$  small enough we reach a contradiction in \eqref{3.33g}. This proves that the constant $1/4$ in \eqref{1.11} is sharp. In the scalar case a similar   argument was used in \cite{bt}.
This completes the proof of Theorem~\ref{theorem1}.

\begin{remark}\label{remark}{\rm As mentioned in the introduction taking the boundary matrix $A$ invertible in the boundary condition \eqref{1.5} amounts to exclude Dirichlet boundary conditions in the diagonal representation where the boundary matrices are given by $\tilde{A}, \tilde{B}.$  Formally the purely Dirichlet boundary condition $\psi(0)=0$ corresponds to taking $ B\to \infty$ in which case the reverse Lieb--Thirring inequality amounts to $\sum_{j}\sqrt{|\lambda_j|} > -\infty,$ which is, of course, trivially always satisfied. Moreover, as is well known,  in the case of the purely Dirichlet  boundary condition in the small coupling constant limit there are no bound states. For the reader's convenience we give the simple proof of this fact assuming that the potential
 belongs to the Faddeev class $L^1_1(\mathbb R^+).$ 

Consider the matrix Schr\"odinger operator with purely Dirichlet boundary condition
$H_{0,I}(\beta \,Q),$ where the coupling constant $\beta$ is a real number and the selfadjoint matrix potential $Q\in L^1_1(\mathbb R^+).$ Let us denote by $H^1_0(\mathbb R^+, \mathbb C^n)$ the completion of $C^\infty_0(\mathbb R^+, \mathbb C^n)$ in the norm of $H^1(\mathbb R^+, \mathbb C^n)$.  The quadratic form of  $H_{0,I}(\beta\, Q)$
 is given by
 \beq\label{3.34b}
q_{0, I, \beta\, Q}(\phi, \psi):= \sum_{i=1}^n \left(\phi_j', \psi_j '\right)+ \beta (Q\phi, \psi),\qquad
\phi, \psi \in H^1_0(\mathbb R^+, \mathbb C^n). 
\ene
 As for $\phi \in H^1_0(\mathbb R^+, \mathbb C^n)$ we have $\phi(0)=0,$
 $$
 \phi_j(x)= \int_0^x \phi_j(y)'\, dy,\qquad j=1,\dots, n.
 $$\
 Then, by Schwarz's inequality
 \beq\label{3.35bb}
 |\phi_j(x)| \leq  \|\phi_j'\|_{ L^2(\mathbb R^+)}\, \sqrt{x}, \qquad j=1,\dots, n.
 \ene
 Moreover, by \eqref{3.34b} and \eqref{3.35bb}
 \beq\label{3.36b}
 q_{0, I, \beta Q}(\phi, \phi)\geq   \left[ \sum_{i=1}^n \left(\phi_j', \phi_j '\right)\right] \left(1-|\beta|  \int_0^\infty\, x\,|V(x)|\,dx \right), \qquad
\phi \in H^1_0(\mathbb R^+, \mathbb C^n). 
\ene
It follows that 
\beq\label{3.37b}
q_{0, I, \beta\, Q}(\phi, \phi)\geq  0,
\ene
if
\beq\label{3.38b}
1-|\beta|  \int_0^\infty\, x\, |V(x)|\,dx  \geq 0.
\ene
Finally  $H_{0,I}(\beta Q)$ has no negative eigenvalues if \eqref{3.38b} holds. 

In conclusion, we have excluded Dirichlet boundary conditions to obtain a meaningful reverse Lieb--Thirring inequality.
}       
\end{remark}


\begin{thebibliography}{99}
\bibitem{AW2021}
T. Aktosun and R. Weder, \textit{Direct and Inverse Scattering for the Matrix Schr\"odinger Equation}, Springer, Switzerland, 2021.

\bibitem{AW2024}T. Aktosun and R. Weder, \textit{The transformations to remove or add bound states for the half-line matrix Schr\"odinger operator}, arXiv: 2402.12136 [math-ph] (2024).

\bibitem{bfs} S. Bachman, R. Froese, and S. Schraven, \textit{Two-sided Lieb--Thirring bounds},  J. Spectr. Theory {\bf 13 }, 1445-1472 (2023).      arXiv:2403.19023 [math-ph]. 



\bibitem{BG2003} A. Ben-Israel and T. N. E. Greville, \textit{Generalized inverses Theory and Applications}, 2nd ed., Springer, New York, 2003.

\bibitem{bk} G. Berkolaiko and P. Kuchment, \textit{Introduction to Quantum Graphs}, AMS, Providence, RI, 2013.

\bibitem{bt} A. Boumenir and V. K. Tuan, \textit{A trace formula and Schminke inequality on the half-line}, Proc. Amer. Math. Soc. {\bf 137},  1039-1049 (2009). 

\bibitem{CM2009} S. L. Campbell and C. D. Meyer, \textit{Generalized Inverses of Linear Transformations}, SIAM, Philadelphia, 2009.

\bibitem{CS1989} K. Chadan and P. C. Sabatier, \textit{Inverse Problems in Quantum Scattering Theory,} 2nd ed., Springer, New York, 1989.

\bibitem{dr} D. Damanik and C. Remling, \textit{Schr\"odinger operators with many bound states}, Duke. Math. J.  {\bf 136}, 51-80 (2007).
\bibitem{ef} T. Ekholm and R. L. Frank, \textit{Lieb--Thirring inequalities on the half-line with critical exponent,}
J. Eur. Math. Soc. {\bf 10}, 739-755 (2008).

\bibitem{elu} P. Exner, A. Laptev, and M. Usman, \textit{On some sharp spectral inequalities for  Schr\"odinger operator on the semi axis}, Comm. Math. Phys {\bf 326}, 531-541 (2014).

\bibitem{flw}  R. L. Frank, A. Laptev, and T. Weidl, \textit{Schr\"odinger Operators: Eigenvalues and Lieb--Thirring Inequalities}, Cambridge University Press, Cambridge, 2023. 



   \bibitem{gl} I. M. Gel'fand  and B. M. Levitan, \textit{ On the determination of a differential equation from its spectral function,} Izv. Akad. Nauk SSSR Ser. Mat. {\bf 15}, 309-360 (1951), in Russian [Am. Math. Soc. Transl. (ser. 2) {\bf 1},  253-304 (1951)  English translation].
   
\bibitem{glaser} V. Glaser, H. Grosse, and A. Martin, \textit{Bounds on the number of eigenvalues of the Schr\"odinger operator},  Comm. Math. Phys. {\bf 59}, 197-212 (1978).

\bibitem{hunder} D. Hundertmark, E. H. Lieb, and L. E. Thomas, \textit{A sharp bound for an eigenvalue moment of the one-dimensional Schr\"odinger operator}, Adv. Theor. Math. Phys. {\bf 2}, 719-731 (1998).

\bibitem{kato} T. Kato, \textit{Perturbation Theory for Linear Operators}, 2nd ed., Springer, Berlin, 1976.

\bibitem{ku} P. Kurasov, \textit{Spectral Geometry of Graphs}, Birk\"auser/Springer, Berlin, 2024.

\bibitem{ll} L. D. Landau and E. M. Lifschitz,\textit{Quantum Mechanics, Non Relativistic Theory,} 3rd ed., Pergamon Press, New York, 1989.

\bibitem{lev} B. N. Levitan, \textit{Inverse Sturm--Liouville Problems}, VNU Science Press, Utrecht, 1987.
 
\bibitem{lg} B. M. Levitan and M. G. Gasymov, \textit{ Determination of a differential operator by two of its  spectra}, Russian Math. Surveys {\bf 19},  1-63 (1964).

\bibitem{lieb-thi} E. H.  Lieb and W. E. Thirring, \textit{Inequalities for the moments of the eigenvalues of the Schr\"odinger Hamiltonian and their relations to Sobolev inequalities.} Pages 269-303 of:  \textit{ Studies in Mathematical Physics (Essays in Honor of Valentin Bargmann)}. Editors: E. H. Lieb, A.S. Whightmann, and B. Simon, Princeton University Press, Princeton, NJ, 1976.

\bibitem{mar} V. A. Marchenko, \textit{Sturm--Liouville Operators and Applications, Revised ed.},  AMS Chelsea, Providence, RI, 2011.  

\bibitem{rs4} M. Reed and B. Simon, \textit{Methods of Modern Mathematical Physics IV  Analysis of Operators}, Academic Press, New York, 1978.

\bibitem{schimm} L. Schimmer, \textit{Improved spectral inequalities for Schr\"odinger operators on the semi-axis}, J. Spectr. Theory {\bf 13}, 47-62 (2023).



\bibitem{schmi} U. -W. Schmincke, \textit{On Schr\"odinger's factorization method for Sturm-Liouville operators}, Proc. Royal Soc. Edinburgh  {\bf 80 A},  67-84 (1978). 


\bibitem{wei} T. Weidl, \textit{On the Lieb--Thirring constant $L_{\gamma,1}$ for $ \gamma \geq 1/2$}, Comm. Math.
Phys. {\bf 178}, 135-146 (1996). 





\end{thebibliography}
\end{document}